# Design, implementation and deployment of the Saclay muon reconstruction algorithms (Muonbox/y) in the Athena software framework of the ATLAS experiment

A. Formica
*DAPNIA/SEDI, CEA/Saclay, 91191 Gif-sur-Yvette CEDEX, France*

This paper gives an overview of a reconstruction algorithm for muon events in ATLAS experiment at CERN. After a short introduction on ATLAS Muon Spectrometer, we will describe the procedure performed by the algorithms Muonbox and Muonboy (last version) in order to achieve correctly the reconstruction task. These algorithms have been developed in Fortran language and are working in the official C++ framework Athena, as well as in stand alone mode. A description of the interaction between Muonboy and Athena will be given, together with the reconstruction performances (efficiency and momentum resolution) obtained with MonteCarlo data.

## 1. INTRODUCTION

This paper summarizes the results collected in ATLAS Technical Design Reports (TDR) [1] [2] and in the recent analysis of Data Challenge 1 (DC1).

## 2. MUON RECONSTRUCTION IN ATLAS

A robust muon identification and high precision momentum measurement is crucial to fully exploit the physics potential of the LHC. The muon energy of physics interest ranges over a large interval from a few GeV, *e.g.* for *B*-physics studies, up to a few TeV, where one might expect the presence of new physics. From the point of view of muon detection, the ATLAS apparatus is characterised by two high precision tracking systems, namely the Inner Detector and the Muon System, and a thick calorimeter that ensures safe hadron filtering with high purity for muons of momentum above 3 GeV. The Muon System is designed to reach high reconstruction efficiency and a momentum resolution at a few % level over a large fraction of the $(\eta,\varphi)$ space covered by the detector, in a wide range of muon energies ($p_T$ from 10 GeV up to a few TeV). The Muon System is composed by: 1) the Muon Drift Tubes (MDT), that are arranged in chambers in three (four) layers around the beam axis: the inner, the middle, and the outer layer. Each MDT chamber contains two superlayers which each consist of 3 or 4 tube layers. An MDT chamber is able to measure a 'local straight track segment' or track vector. The muon trigger chambers are Resistive Plate Chambers (RPC) in the barrel and Thin Gap Chambers (TGC) in the end-caps and are physically connected to the MDT chambers. A package of MDT and RPC/TGP chambers is called a station. The whole of the spectrometer is in a toroidal magnetic field (see Figure 1).

Muon reconstruction in ATLAS experiment is a very challenging task, due to several effects: the large muon system size (44m length and 22m Ø) implies higher extrapolation uncertainties; the open air-core toroid that leads to a rather inhomogeneous magnetic field affects the measurement for events with low $p_T$ momentum, and finally the precision chambers that can measure only the coordinate giving the bending direction with an accuracy of 80μm, while the second coordinate is given by trigger chambers with an accuracy of 1cm. High level of physics background is also expected.

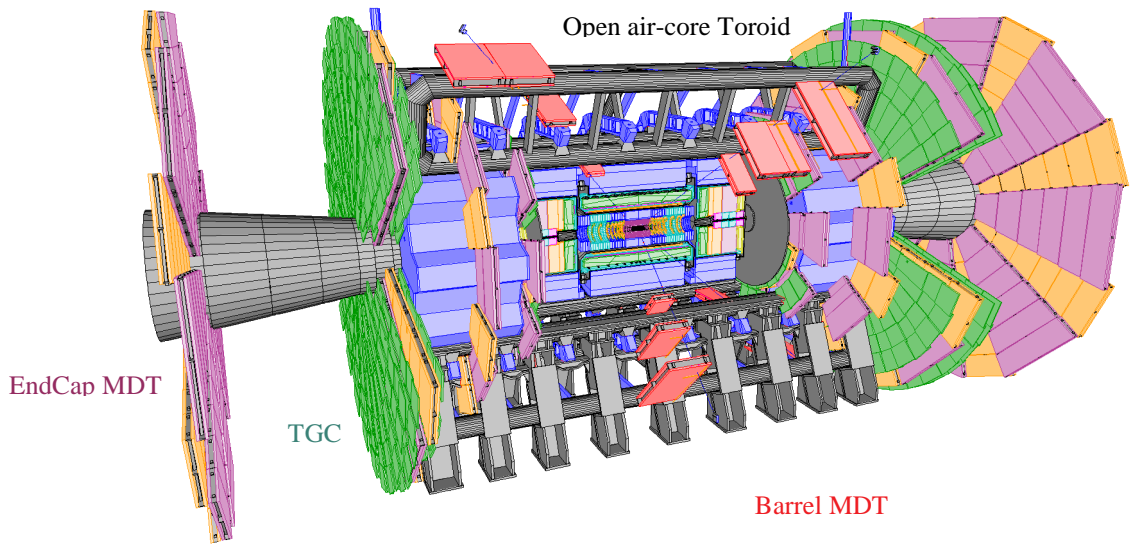

Figure 1: ATLAS detection of H→4μ simulated event. The MDT chambers traversed by the muons are shown

## 3. MUONBOX (MUONBOY) ALGORITHMS

These algorithms have been developed in F77 (Muonbox) and F90 (Muonboy). The usage of the latter language allows many improvements in the code, since F90 contains advanced features like dynamic memory management, recursion and polymorphism

### 3.1. Strategy for pattern recognition

The strategy of the pattern recognition algorithm, described below, can be summarized in four main points: 1) identifications of 'regions of activity' (ROA) in the muon system, through the RPC/TGC systems; 2) reconstruction of 'local straight track segments' in each muon station of these regions of activity; 3) combination of tracks segments of different muon stations to form muon track candidates; 4) global track fit of the muon track candidates through the full system.

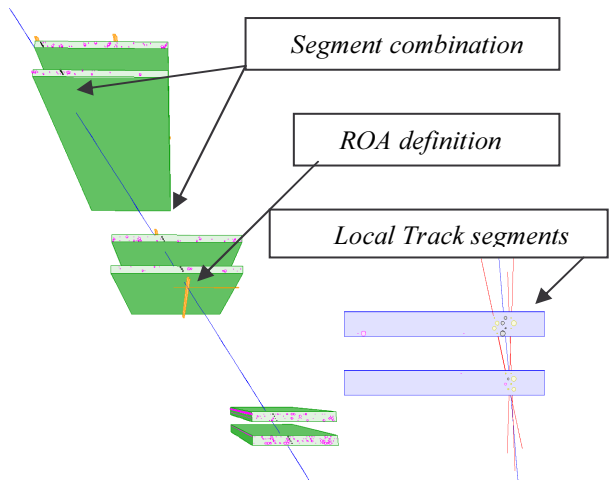

Figure 2: pattern recognition strategy

In the first step, ROAs in the ($\eta$,$\varphi$) space are identified using information from the trigger chambers. The size of these regions is roughly d$\eta \times$ d$\varphi = 0.4 \times 0.4$ and they are centered where there exists at least one RPC/TGC hit in both coordinates (among up to seven layers of RPC/TGC). Then, straight track segments are first reconstructed individually, in the bending plane, in each muon station intersecting with these ROAs. We will enter in the detail of segment combination later in this paper (paragraph **3.3**).

### 3.2. Input Parameters for Reconstruction

The needed input parameters for the algorithm are:
1. **drift time** (dt) measurement in MDT tubes and ROA hits coming from trigger chambers. The drift time is corrected accounting for Lorentz angle (magnetic field effect), tube length (signal propagation delay), and time of flight;
2. the **RT** relations (time vs distance), that are measured using real tracks;

3. an accurate **geometry description** of the whole detector (AMDB database), needed to correctly take into account the traversed material during the tracking procedure;
4. **alignment corrections** to nominal chamber positions are also important for tracking the muons;
5. the **magnetic field** (figure 3) has to be known with a precision of 20-30 gauss ($10^{-3}$) that is required in the fit procedure. The magnetic field will be measured by sensors installed in the detector. A service has been developed in order to provide magnetic field values for every given space point:

$$f(x, y, z) \rightarrow (\vec{B}, \nabla \vec{B})$$

Using these parameters and the previously described strategy for the pattern recognition, we are able to reconstruct tracks inside the spectrometer

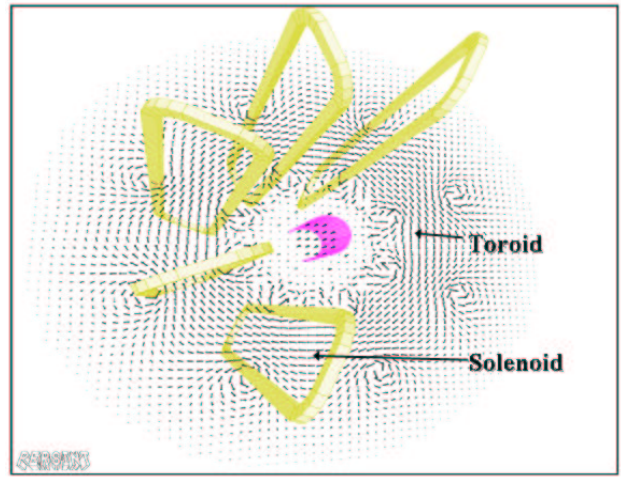

### 3.3. Building Tracks

Starting from straight track segments from each muon station we can combine them and perform track reconstruction in the whole spectrometer. During this procedure, the algorithm takes into account for $\delta$ rays and tube inefficiencies that can degrade the $\chi^2$ of the track. The position and direction of track segments found in the outer and middle stations of the muon spectrometer allow a first rough estimate of the momentum of corresponding candidate muons. Each of these strict track segments is then extrapolated to the first middle or outer (or even inner when middle or outer is missing) station found using tracking in the magnetic field. A momentum scan around the first estimate is used searching for best matching between tracks. Fit on the candidate track is made performing full tracking at each step of the minimisation procedure. In this way the algorithm can fine tune position, direction, and momentum of the track, by taking into account for inefficient tubes (due to dead time in the detector), and for missing hits along the track trajectory.

Finally, a last global re-fit is performed, starting from the best result of the previous fits, but using this time directly the raw information available information (i.e. the drift time values and hit strips instead of the pre-reconstructed straight track segments). Eventually, the selection of reconstructed muons is made according to the value of the $\chi^2$ of this last global fit. Those single hits that are spoiled by δ rays and neutron background are cleaned up during this final fit. Multiple scattering and energy loss in the spectrometer matter are taken into account as well, using the geometry description.

### 3.4. Tracking Back to the Beam

Another relevant task of the reconstruction is to perform backtracking from Muon System down to beam region. This procedure requires an accurate knowledge of the amount and the nature of the material traversed by muon trajectory in such a way to correctly account for energy losses of the muons along the track. Muon momentum is corrected using an energy loss parameterization. A second method that will be implemented in future can use the calorimeter energy measured in the traversed cells. This second method is foreseen however only for high $p_T$ muons, having higher probability of a catastrophic energy loss. The reconstruction program is able to predict with good accuracy (see figure 4) the energy loss over a wide range of muon momenta (5-500 GeV). A complete set of output parameters is provided by Muonbox and Muonboy algorithms: the track parameters and the covariance matrix of the fit. Track parameters are given in different positions of the detector: the entrance of the spectrometer, the entrance of the calorimeters, and at the perigee level (i.e. the closest distance of approach with the beam axis). The covariance matrix represents the errors on tracking procedure, and takes into account for multiple scattering and fluctuations in energy losses.

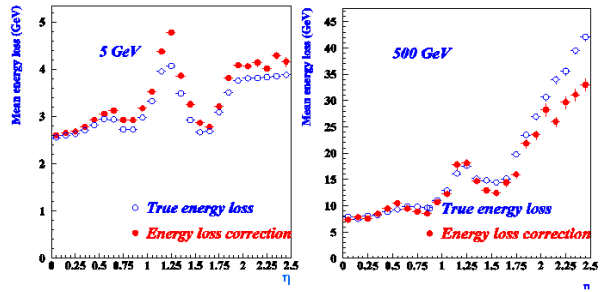

### 3.5. Combination with Inner Detector

Muon tracks are also reconstructed in the Inner Detector. The combination of measurements made in the Muon System with the ones from the Inner Detector improves the momentum resolution in the momentum range 6 GeV < $p_T$ < 100 GeV. The matching of the muon track reconstructed independently in the Inner Detector and in the Muon System allows the rejection of muons from secondary interactions as well as the ones from π/K decays in flight.

In order to combine the tracks reconstructed in the Inner Detector and the Muon System, Muonbox/y apply a strategy based on the statistical combination of the two independent measurements using the parameters of the reconstructed tracks and their covariance matrices. We will not enter here in a detailed description of this method, which can be found in [2].

### 4. ALGORITHM PERFORMANCES

In figures 5 and 6 we show the reconstruction efficiency and the resolution for the Muon spectrometer alone as well as for the combined reconstruction with the Inner Detector. The plots reproduce the TDR result, and are obtained running Muonbox algorithm inside Athena framework, on a data sample coming from DC1 (Data Challenge 1).

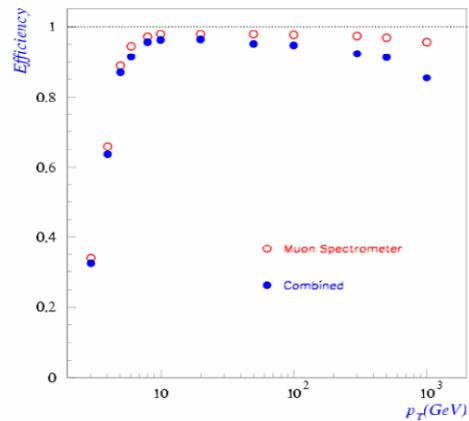

Figure 5: Muonbox efficiency on DC1 data

The reconstruction efficiency is affected by the energy deposit of the muon in the detector before the Muon System for low $p_T$ muons.

The main contribution to reconstruction resolution are coming from the Muon System alignment (30 μm for Barrel, 50μm for End-Cap) and from the space resolution of the precision chambers at high momentum ($p_T$ > 200 GeV), from energy loss fluctuations at $p_T$ < 200 GeV, and from multiple scattering fluctuations that are almost constant in $p_T$.

Figure 6: Muonbox: $p_T$ resolution on DC1 data

The time performance (time per event) of the reconstruction is summarized in figure 7, for three different versions of the algorithm: Muonbox in blue, Muonboy in red and an accelerated version in violet. The measurement were performed on a DEC alpha station (700 MHz), running the algorithm in a stand alone mode.

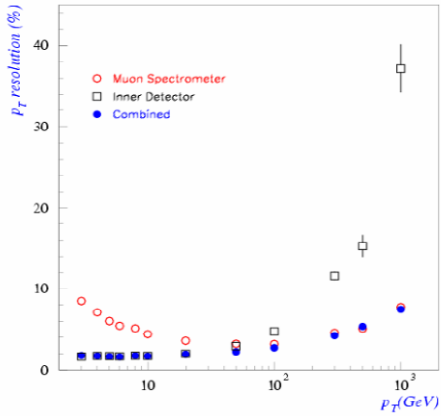

Figure 6: Muonbox: $p_T$ resolution on DC1 data

The time performance (time per event) of the reconstruction is summarized in figure 7, for three different versions of the algorithm: Muonbox in blue, Muonboy in red and an accelerated version in violet. The measurement were performed on a DEC alpha station (700 MHz), running the algorithm in a stand alone mode.

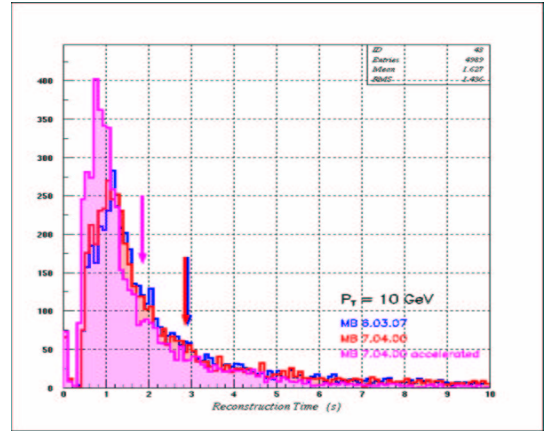

Figure 7: Muonbox/y time performances

The accelerated version of the algorithm is foreseen to improve the time per event performances in order to have the possibility of using this reconstruction for high level trigger purposes.

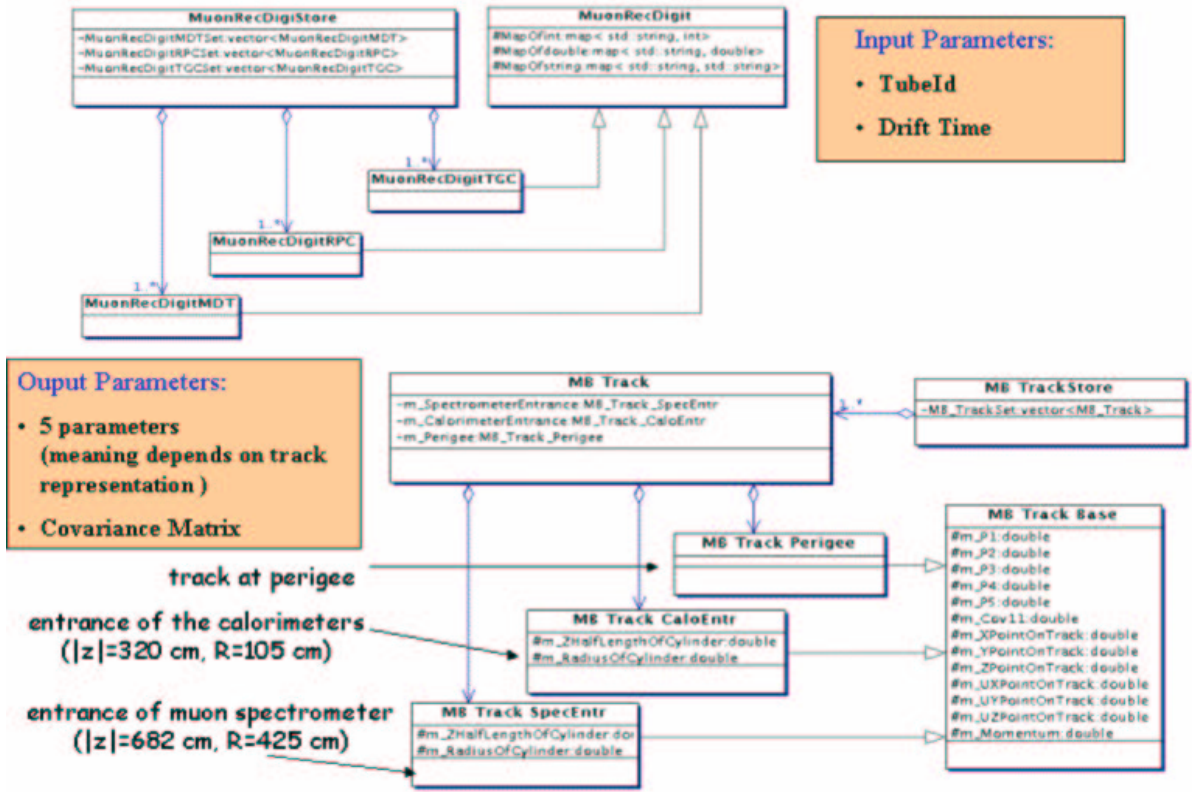

Figure 8: Diagrams of the interface between Muonboy and Athena framework

## 5. USING MUONBOX/Y

### 5.1. Athena framework

Athena is the official C++ framework that provides services and algorithms for physics analysis in ATLAS.

Algorithms inside Athena can access the hits information needed for the reconstruction from the Transient Data Store, and put the reconstruction output (track parameters and covariance matrix) in the same place, from where other users can access it for further analysis. We recall here that the output parameters are given in different position of the detector, so to allow combination of muon spectrometer tracks with the Inner Detector tracks (see figure 8). An interface has been developed in order to run Muonbox/y algorithms in Athena. This interface takes care of the information transmission from the C++ world to the reconstruction algorithms.

## 5.2. Stand Alone Mode

The algorithms can run in stand alone mode by reading all needed input from external files (geometry and magnetic field map as well). This feature has been very useful in the debugging of the algorithm itself, and for a fast feedback in the analysis of TestBeam data in H8 at CERN, during 2002 and 2003 years. For its modularity, this reconstruction is used also in an event display (see figure 1) that gives a visual representation of the geometry of the detector, and can simulate and reconstruct muon events at once.

## Acknowledgments

The authors wish to thank Saclay Muon Software Group for the help and guidance in preparing this document.